\documentclass[aps,prd,noshowpacs,twocolumn,footinbib,floatfix,citeautoscript,preprintnumbers]{revtex4}
\usepackage[english]{babel}
\usepackage{amssymb}
\usepackage{amsfonts}
\usepackage{amsmath,amssymb}
\usepackage{eucal}
\usepackage{graphicx}
\usepackage{epsfig,color,subfigure}
\usepackage{slashed}
\usepackage[bookmarks=true,colorlinks,linkcolor=red,urlcolor=blue,citecolor=blue]{hyperref}
\newcommand{\be}{\begin{equation}} \newcommand{\ee}{\end{equation}}
\newcommand{\bea}{\begin{eqnarray}} \newcommand{\eea}{\end{eqnarray}}
\newcommand{\beann}{\begin{eqnarray*}}  \newcommand{\eeann}{\end{eqnarray*}}
\newcommand{\bfig}{\begin{figure}} \newcommand{\efig}{\end{figure}}
\newcommand{\ba}{\begin{array}} \newcommand{\ea}{\end{array}}
\newcommand{\bcen}{\begin{center}} \newcommand{\ecen}{\end{center}}
\newcommand{\btab}{\begin{tabular}} \newcommand{\etab}{\end{tabular}}

\newtheorem{Proposition}{Proposition}[section]

\newtheorem{Theorem}{Theorem}[section]
\newtheorem{Lemma}{Lemma}[section]
\newtheorem{Corrolary}{Corrolary}[section]

\newcommand{\bp}{\begin{Proposition}}	\newcommand{\ep}{\end{Proposition}}
\newcommand{\bt}{\begin{Theorem}}	\newcommand{\et}{\end{Theorem}}
\newcommand{\bl}{\begin{Lemma}}		\newcommand{\el}{\end{Lemma}}
\newcommand{\bc}{\begin{Corrolary}}	\newcommand{\ec}{\end{Corrolary}}
\begin{document}
\title{The holographic Weyl semi-metal}

\author{Karl Landsteiner}\email{Karl.Landsteiner@csic.es}
\author{Yan Liu}\email{yan.liu@csic.es}
\affiliation{Instituto de F\'{\i}sica Te\'orica UAM/CSIC, C/ Nicol\'as Cabrera
13-15,\\
Universidad Aut\'onoma de Madrid, Cantoblanco, 28049 Madrid, Spain}

\begin{abstract}
We present a holographic model of a Weyl semi-metal. We show that upon varying a mass parameter
the model undergoes a quantum phase transition from a topologically non-trivial state to a trivial one.
The order parameter for this phase transition is the anomalous Hall effect (AHE). We give an interpretation
of the results in terms of a holographic RG flow and compare to a weakly coupled field theoretical model.
Since there are no quasiparticle excitations in the strongly coupled holographic model the topological phase 
can not be bound to notions of topology in momentum space. 

\end{abstract}

\preprint{IFT-UAM/CSIC-15-046}
\maketitle
%

\begin{figure}[h!]
\includegraphics[width=0.3\textwidth]{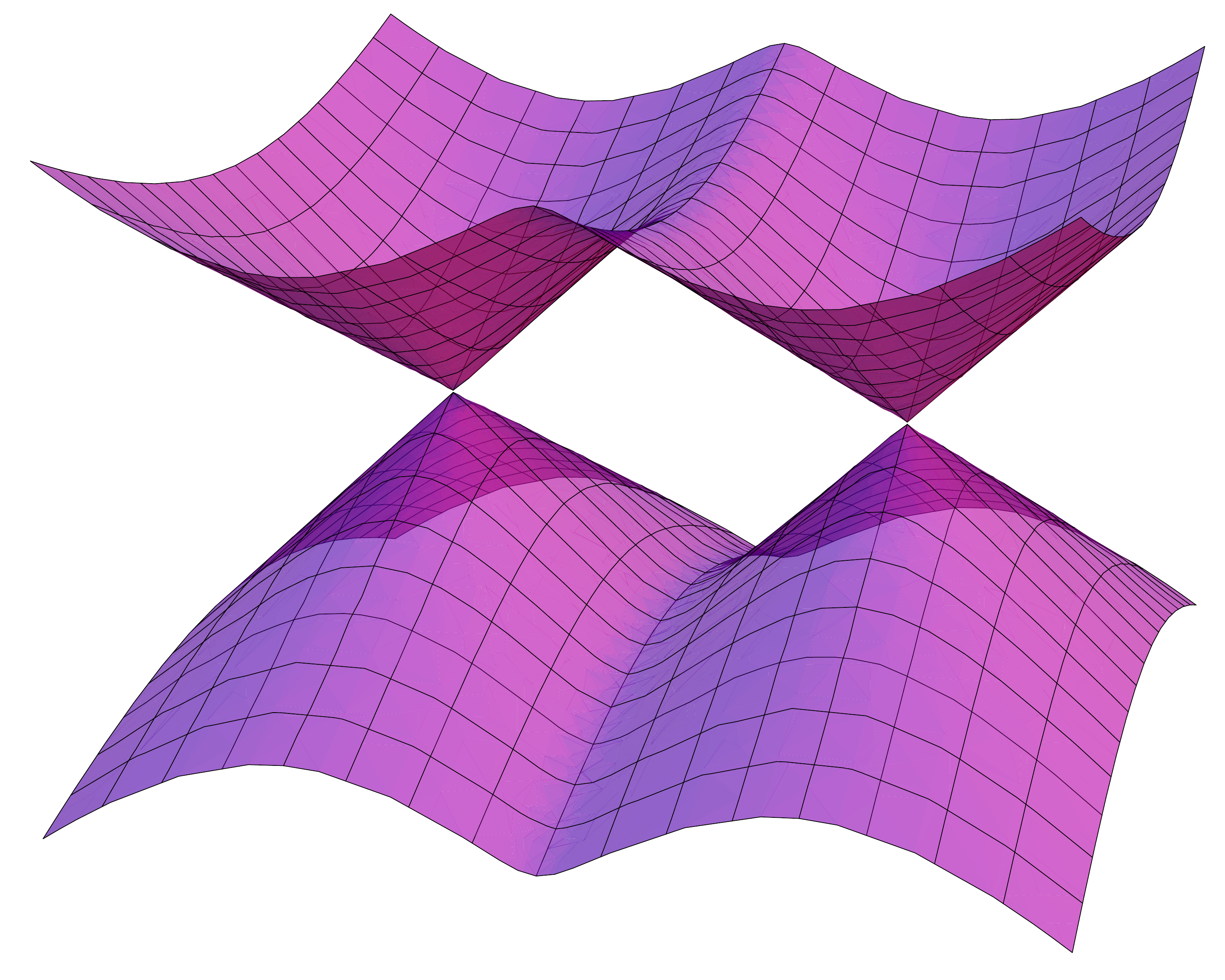}
\caption{\small 
Locally around two band touching points separated by a vector ${\bf b}_\text{eff}$ the dispersion of the quasiparticles
forms two Weyl cones of left- and right handed chirality. Both nodes lie precisely at the Fermi energy $\epsilon_F$.}
\label{fig:Weylspectrum}
\end{figure}

Weyl semi-metals are an exciting new class of materials with exotic electronic transport properties (for reviews see \cite{Hosur:2013kxa,wsmreview2}).
As a semi-metal their Fermi surface consists of isolated points. The electronic quasiparticle excitations
around these points can be described by pairs of left- and right-handed Weyl spinors. The crucial property
is that these singular points in the band structure of a crystal are separated by a (spatial) vector
in momentum space. Because of topological constraints there always have to appear pairs of left-
and right-handed Weyl spinors \cite{Nielsen:1983rb,volovik}, see Fig. \ref{fig:Weylspectrum}.

\begin{figure}
 \includegraphics[width=.23\textwidth]{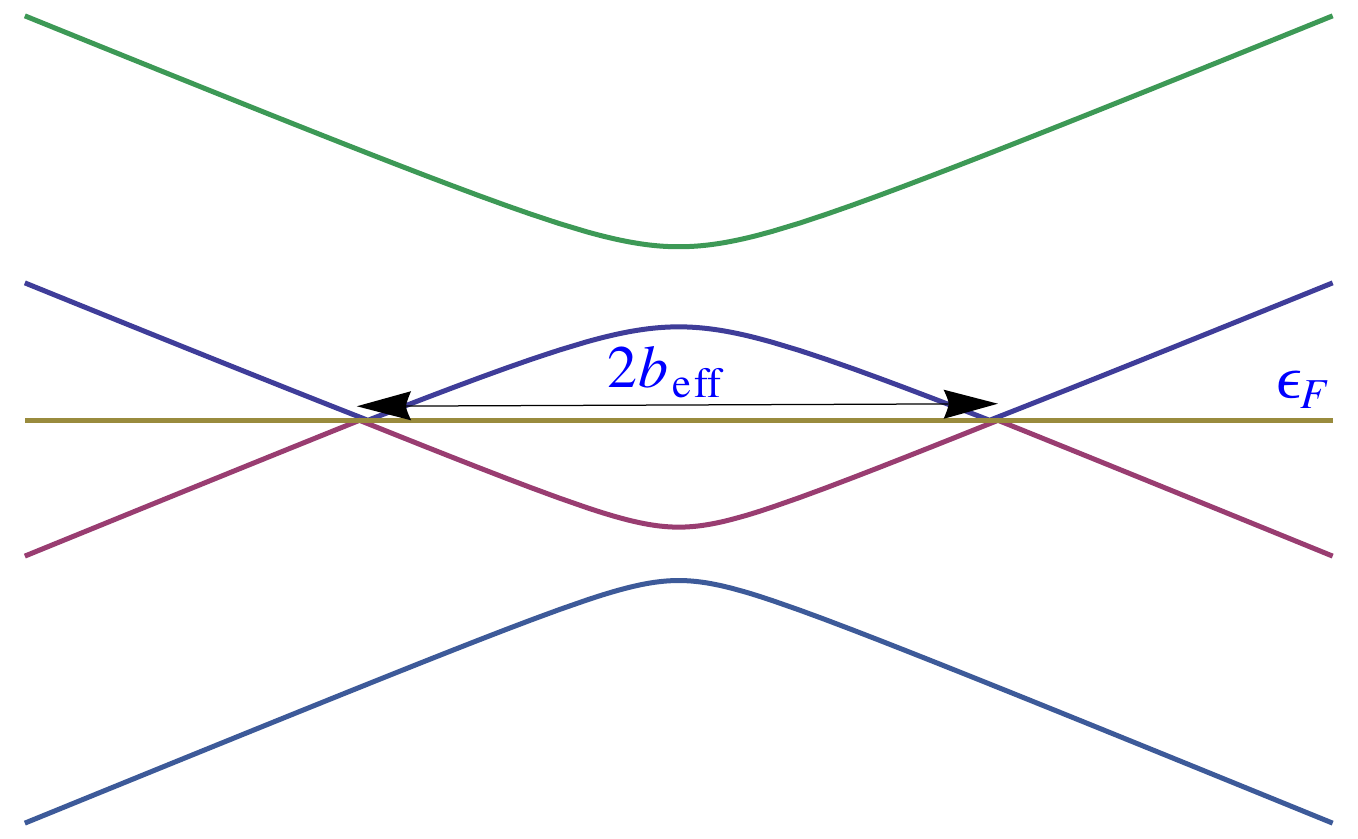}
 \includegraphics[width=.23\textwidth]{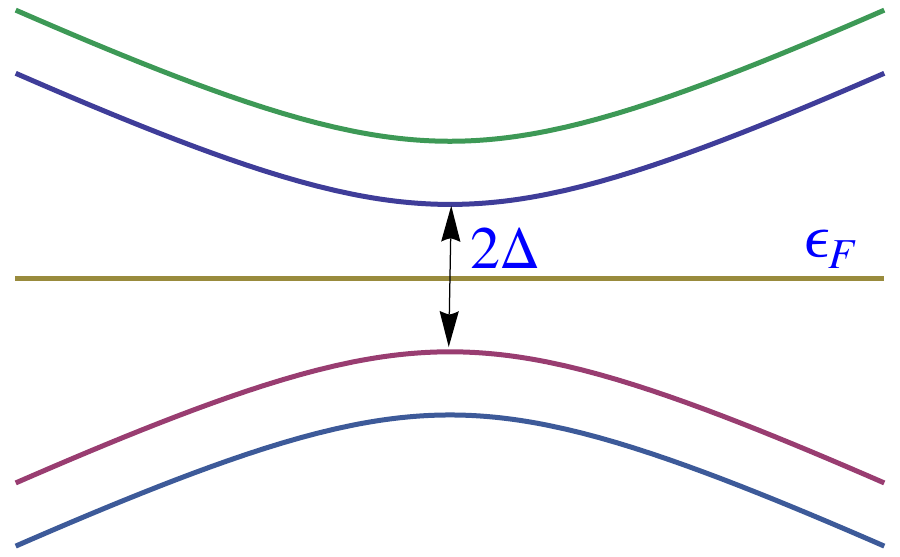}
 \caption{\small Left panel: For $b^2>M^2$ there are two Weyl nodes in the spectrum. They are separated by the distance $2\sqrt{b^2-M^2}$ in momentum space.
Right panel: For $b^2<M^2$ the system is gapped with gap $2\Delta = 2\sqrt{M^2-b^2}$.}
\label{fig:topphase}
\end{figure}

The fact that the left- and right-handed Weyl nodes are separated by a spatial vector means that time reversal 
symmetry is broken. It is common to make the simplifying assumption that the Weyl cones are rotationally 
symmetric and that their opening angles are equal.
Under these assumptions a quantum field theoretical model can be constructed. It takes the form of a ``Lorentz breaking'' Dirac
system with Lagrangian \cite{Colladay:1998fq}
\begin{equation}
\label{eq:lagrangian}
\mathcal{L} = 
\bar\Psi \left( i\slashed\partial - e \slashed{A} - \gamma_z\gamma_5 b + M \right)\Psi\,.
\end{equation}
At $M=0$ the parameter $b$ separates the left- and right-handed spinors by a distance of $2b$ in momentum space along the $z$-direction. On the other hand for $b=0$ but $M\neq 0$ one is dealing with massive fermions. For arbitrary values of $b$ and $M$ the spectrum can be easily obtained and is sketched in Fig. \ref{fig:topphase}.
As long as $|b|>|M|$ the spectrum is ungapped. It is characterized by band inversion and at the crossing points the wave function is well-described by Weyl fermions. The separation of the Weyl cones is given by $2\sqrt{b^2-M^2}$. In this situation the quantum field theoretical model can be further reduced to an effective model with Lagrangian
\begin{equation}\label{eq:lowenergychiral}
 \mathcal{L} = \bar \psi \left( i \slashed{\partial} - e \slashed{A} - \gamma_z\gamma_5 b_\text{eff} \right)\psi\,
\end{equation}
with $b_\text{eff} = \sqrt{b^2-M^2}$.

If $|b|<|M|$ then the system is gapped and the low energy description is simply one of a massive Dirac fermion
\begin{equation}\label{eq:lowenergygapped}
 \mathcal{L} = \bar \psi \left( i \slashed{\partial} - e \slashed{A} + \Delta \right)\psi\,,
\end{equation}
and $\Delta = \sqrt{M^2-b^2}$.
Accordingly the system undergoes a quantum phase transition from the topologically non-trivial Weyl semi-metal phase to a trivial insulating phase.

Let us now give a quick derivation of the anomalous Hall effect using the effective field model (\ref{eq:lowenergychiral}). In this model the low energy
axial symmetry is unbroken up to the axial anomaly. Therefore a redefinition of the low energy spinor $\psi \rightarrow \text{exp}(i\gamma_5 \theta)\psi$ will
induce the anomaly term upon integrating over the redefined spinors. Choosing $\theta = 
\bf{b}_\text{eff}\cdot\bf{x}$ we find the anomaly term $\Gamma_\text{anom} =
e^2/(16\pi^2) \int d^4x\, ({\bf{b}}_\text{eff}\cdot {\bf{x}}) \epsilon^{\mu\nu\rho\lambda} F_{\mu\nu} F_{\rho\lambda}
$. The current can then be computed as variation
with respect to the gauge field and we find the anomalous Hall effect \cite{Yang,Xu:2011dn,burkovbalents,Grushin:2012mt,Goswami:2012db,Zyuzin:2012tv,Chen:2013mea,vazifeh}
\begin{equation}\label{eq:ahe}
 {\bf{J}} = \frac{e^2}{2\pi^2} \bf{b}_\text{eff} \times \bf{E}\,.
\end{equation}

On the other hand we can directly work with the ``high energy'' model (\ref{eq:lagrangian}). In this case the axial symmetry has both, an anomaly and
a tree level breaking term
\begin{equation}
 \partial_\mu J^\mu_5 = \frac{1}{16\pi^2} \epsilon^{\mu\nu\rho\lambda} F_{\mu\nu} F_{\rho\lambda} + 2 M \bar\Psi\gamma_5 \Psi\,.
\end{equation}
The anomalous Hall effect can now be computed as a one-loop contribution to the polarization tensor. This calculation has a long history and is plagued
by regularization ambiguities \cite{Jackiw:1999qq}.
As we will see in the holographic model allows to resolve the ambiguities in a unique form.

The topological property in the Weyl semi-metal phase is intimately related to the fact that the wave function of a Weyl spinor can be understood as a monopole of the Berry curvature
in momentum space, with the left-handed Weyl fermion having monopole charge $+1$ and the right-handed one having monopole charge $-1$ \cite{volovik}. In an inherently strongly coupled system the status of the single
particle wave function is not a priori clear. Moreover at strong coupling the concept of quasiparticle might not even be applicable. The question arises then 
if it is possible to construct a model at strong coupling that has the essential physical properties of a Weyl semi-metal. In particular does there exist a strongly
coupled model in which the anomalous Hall effect and a quantum phase transition to a topological trivial phase persists even in the absence of the notion of
singularities in the dispersion relations of fermionic quasiparticles? String theory inspired holography based on the AdS/CFT correspondence has arisen in the last
few years as a singular and useful tool to address such questions. Holography has indeed already proved to be extremely useful for the
understanding of  transport properties of relativistic systems. In particular the modern understanding of anomaly related transport phenomena such as the
chiral magnetic \cite{Fukushima:2008xe} and chiral vortical effects \cite{Banerjee:2008th,Erdmenger:2008rm,Son:2009tf, Landsteiner:2011iq} is based to a considerable part on research
using holographic models \footnote{Previous holographic treatments of Weyl semimetals \cite{Jacobs:2014nia,Gursoy:2012ie} differ from our approach in that they study the holographic fermionic spectral functions.}.

We consider the following holographic action which was studied in \cite{Jimenez-Alba:2015awa} to encode the axial charge dissipation effect  in order to get a finite longitudinal DC magnetoconductivity \cite{Landsteiner:2014vua}
\begin{align}\
  S=&\int d^5x \sqrt{-g}\bigg[\frac{1}{2\kappa^2}\Big(R+12\Big)-\frac{1}{4}F^2-\frac{1}{4}F_5^2+ \nonumber\\
&+\frac{\alpha}{3}\epsilon^{\mu\nu\rho\sigma\tau}A^5_\mu \Big(F^5_{\nu\rho} F^5_{\sigma\tau}+3 F_{\nu\rho} F_{\sigma\tau}\Big)\nonumber\\
&-(D_\mu\Phi)^*(D^\mu\Phi)-m^2\Phi^*\Phi\bigg] \,.\label{eq:holomodel}
\end{align}
As is well known, global symmetries correspond to gauge fields in AdS space. We will need two such gauge field, one representing the the electromagnetic $U(1)$ symmetry.
Its AdS bulk gauge field is denoted by $V_\mu$  and its field strength is $F=dV$. The axial $U(1)$ symmetry is represented by the gauge field $A_\mu^5$ with
field strength $F_5 = dA_5$. It is anomalous and the anomaly is represented in (\ref{eq:holomodel}) by the Chern-Simons part of the action with coupling constant $\alpha$.
Note that the choice of Cher-Simons term is the unique one that makes the electromagnetic symmetry non-anomalous \cite{Rebhan:2009vc}. The axial symmetry is also
broken by the mass term. The mass deformation is introduced via a non-normalizable mode of a scalar field. This scalar field is charged only under the
axial gauge transformation and its covariant derivative is $D_\mu\Phi = (\partial_\mu - i q A^5_\mu )\Phi$.
The scalar bulk mass is chosen such that the dual operator has dimension three, i.e. $m^2L^2 = -3$ where $L$ is the intrinsic length scale of AdS space
\footnote{In the following we will set $L=1$ and also $q=1$.}.
The electromagnetic and axial currents are defined as
\begin{align}\label{eq:consVcur}
J^\mu &= \lim_{r\rightarrow\infty}\sqrt{-g}\Big(F^{\mu r}+4\alpha\epsilon^{r \mu\beta\rho\sigma} A^5_{\beta} F_{\rho\sigma}  \Big)+\text{c.t.} \,,\\
\label{eq:consAcur}
 J^\mu_5 &= \lim_{r\rightarrow\infty}\sqrt{-g}\Big(F_5^{\mu r}+\frac{4\alpha}{3}\epsilon^{r \mu\beta\rho\sigma} A^5_{\beta}F^5_{\rho\sigma}  \Big)+\text{c.t.}\,.
\end{align}
These are the {\em consistent} currents obtained by
the variations of the on-shell boundary action with respect to the electromagnetic and axial boundary values of the field $V_\mu$ and
$A^5_\mu$. It is also common to define {\em covariant} currents \cite{Bardeen:1984pm}. Their AdS definition is given by simply dropping the
Chern-Simons parts in (\ref{eq:consVcur},\ref{eq:consAcur}). The covariant currents are invariant even under the anomalous axial gauge transformations.

We will work in the following in the probe limit in which metric fluctuations are neglected. The metric background is fixed and taken as the AdS Schwarzschild solution
\begin{equation}
 ds^2 = - r^2f(r) dt^2 + \frac{dr^2}{r^2f(r)} + r^2 d{\bf{x}}^2~,~~~f(r)=1-\frac{r_h^4}{r^4}\,.
\end{equation}
Now we introduce the parameters $b$ and $M$ via boundary conditions on the fields $A^5_z$ and $\Phi=\phi(r)$
\begin{equation}
 \lim_{r\rightarrow \infty} A^5_z(r) = b~,~~~~~~\lim_{r\rightarrow \infty} r\phi(r) = M\,.
\end{equation}
The equations of motion for the background solution are
\begin{align}
(A^5_z)''+\bigg(\frac{3}{r}+\frac{f'}{f}\bigg)(A^5_z)'-\frac{2q^2\phi^2}{r^2f}A^5_z&=0\,,\\
\phi''+\bigg(\frac{5}{r}+\frac{f'}{f}\bigg)\phi'-\bigg(\frac{q^2(A^5_z)^2}{r^4f}+\frac{m^2}{r^2f}\bigg)\phi&=0\,.
\end{align}
This system of differential equations can be solved by demanding that $A^5_z$ and $\phi$ are regular at the horizon $r=r_h$.

The Hall conductivity can be computed with the help of the Kubo formula
\begin{equation}\label{eq:Kubo}
 \sigma_{xy} = \lim_{\omega\rightarrow 0}\frac{1}{i\omega} \langle J_x J_y \rangle_R\,.
\end{equation}

In holography the retarded correlation function of two currents can be computed by considering
fluctuations above the background. In particular we need fluctuations of the vector type bulk
gauge field $V_\mu$ in $x$ and $y$ directions, i.e. $\delta V_x= v_x e^{-i\omega t}, \delta V_y=v_y e^{-i\omega t}$. 
The equations for the fluctuations are
\begin{equation}\label{eq:flucs}
v_\pm''+\bigg(\frac{3}{r}+\frac{f'}{f}\bigg)v_\pm'+\frac{\omega^2}{r^4f^2}v_\pm\pm\frac{8\omega\alpha}{r^3f}(A^5_z)'v_\pm=0\,,
\end{equation}
where $v_\pm=v_x\pm i v_y.$

In order to obtain the retarded correlator we impose infalling boundary conditions
at the horizon. Furthermore we only need the leading behavior in an expansion around
zero frequency. A convenient parametrization for the fluctuations is therefore
\begin{equation}
 v_\pm=f^{-\frac{i\omega}{4r_h}}\big(v_\pm^{(0)}+\omega v_\pm^{(1)}+\dots\big)\,.
\end{equation}
 
To zeroth and first order in $\omega$ 
\begin{align}
\label{eq:0th}
{v_\pm^{(0)}}''+\bigg(\frac{3}{r}+\frac{f'}{f}\bigg){v_\pm^{(0)}}'&=0\,,\\
\label{eq:1st}
{v_\pm^{(1)}}''+\bigg(\frac{3}{r}+\frac{f'}{f}\bigg){v_\pm^{(1)}}'&=\bigg[\frac{i}{4r_h}\bigg(\frac{3f'}{rf}+\frac{f''}{f}\bigg)
\mp\frac{8\alpha}{r^3f}(A^5_z)'\bigg]v_\pm^{(0)}\nonumber\\
&+\frac{i}{2r_h}\frac{f'}{f}{v_\pm^{(0)}}'\,.
\end{align}

We impose the regularity condition for $v_\pm^{(0)}$ and $v_\pm^{(1)}$ near horizon.
From (\ref{eq:0th}), we have $v_\pm^{(0)}=c_0$.
From (\ref{eq:1st}), we have
\be
v_\pm^{(1)}=-\int_r^{\infty}dx \frac{c_0}{x^3f}\bigg[
\frac{i x^3 f'}{4r_h}-ir_h\mp 8\alpha \big(A^5_z-A^5_z(r_h)\big)\bigg]\,.
\ee

Thus $G_\pm=\omega\big(\pm 8\alpha (b-A^5_z(r_h))+ir_h\big)$, we have
\begin{equation}\label{eq:Hallcov}
\sigma_{xy}=\frac{G_+-G_-}{2\omega}=8\alpha \big(b-A^5_z(r_h)\big)\,.
\end{equation}
It is important to realize that this is the Hall conductivity in the covariant current.
In order to obtain the total Hall conductivity we have to add the the contribution from the
Chern-Simons term \footnote{We define the $\sigma_{\text{AHE}}$ via  
${\bf{J}} = \sigma_{\text{AHE}} \bf{e}_b \times \bf{E}$ where ${\bf e}_b$ is the unit vector along the Weyl nodes separation direction ${\bf b}$. Thus for free field model (\ref{eq:lagrangian}), we have $\sigma_{\text{AHE}}=\frac{e^2}{2\pi^2}b$. One may identify the anomaly constant $c=8\alpha=\frac{e^2}{2\pi^2}$ \cite{Rebhan:2009vc}.}, 
\begin{equation}
\label{eq:Halltotal}
 \sigma_\text{AHE} = 8\alpha b-\sigma_{xy}= 8\alpha A^5_z(r_h)\,.
\end{equation}
This is the main result.

\begin{figure}[h!]
 \includegraphics[width=0.45\textwidth]{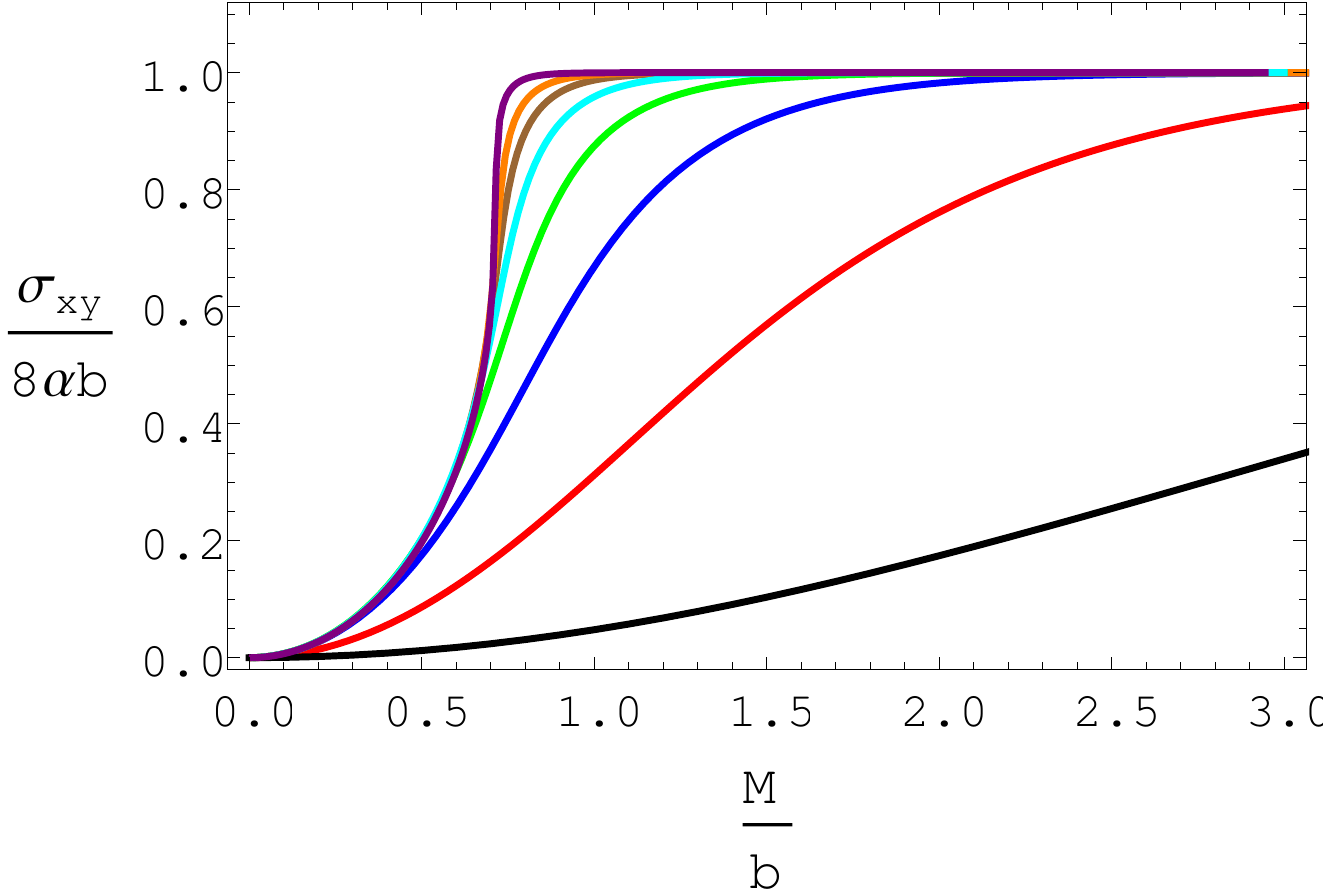}
\caption{\label{fig:ahecov} \small Anomalous Hall conductivity of the covariant current as a function of $M/b$ at different temperatures. From the black curve (bottom) to the purple curve (top) $\pi T/b$ corresponds to $\pi,1, 1/2, 1/3,1/4,1/5,1/6,1/8$ respectively.}
\end{figure}

We solved the equations numerically for different values of the the boundary parameters $M$ and $b$. 
In Fig. \ref{fig:ahecov} we show the anomalous Hall response in the covariant current. 
Because of the underlying conformal symmetry we have fixed $b$ and vary $M/b$ and the temperature $T/b$. The plots show how the response of the covariant
current builds up for high temperature (black curve) to low temperature (purple curve). At low temperature the response
builds up quickly and saturates precisely at the value that cancels the Hall conductivity 
stemming from the Chern-Simons part of the current.

\begin{figure}
 \includegraphics[width=0.45\textwidth]{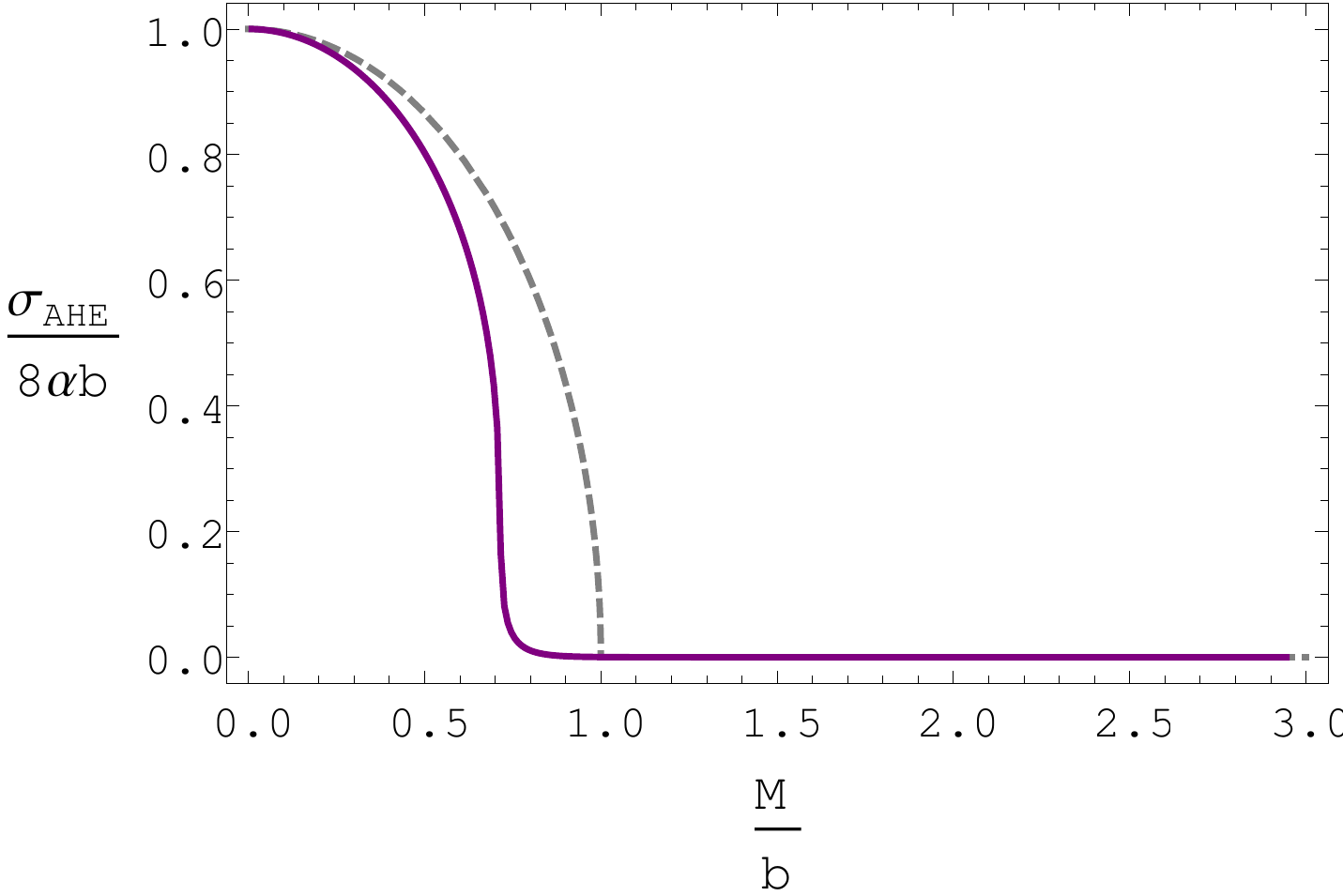}
\caption{\label{fig:Halltotal} \small Anomalous Hall conductivity at low temperature. The solid purple line is for $\pi T/b=0.125$ while the dashed line is the free field result (\ref{eq:ahe}).}
\end{figure}

For low temperatures the change is rather drastic. This is consistent with the idea that the system undergoes a
quantum phase transition from a topologically non-trivial to a trivial state.
In Fig. \ref{fig:Halltotal} we have plotted the anomalous Hall conductivity in the total current at low temperature (we have chosen $\pi T/b=0.125$).
The total Hall conductivity drops off very quickly and basically vanishes at a critical value of the mass $M$. Due to the fact that we work in the probe
limit we can not really reach the zero temperature limit. Accordingly we observe a smooth crossover instead of a sharp (quantum) phase transition.
Nevertheless, already at $\pi T = 0.125b$ the behavior of the anomalous Hall conductivity allows to estimate the critical value of the mass, which
we find to be $M_c \approx 0.7b$. For comparison we also show the anomalous Hall conductivity of the simple free field model (\ref{eq:lagrangian}).

We note that the longitudinal conductivity is independent of $M$ and $b$ with $\sigma_{xx}=\sigma_{yy}=\sigma_{zz}=\pi T$. A linear dependence on the temperature is natural for the ungapped topological phase. It represents the conductivity induced by the thermally activated pairs of charge fermions and anti-fermions (holes). In the topologically trivial phase a linear dependence can be expected only in the case of small gap $\Delta/T <1$. That we observe exact linear dependence even in the trivial phase has to attributed to the probe limit. Indeed it is expected that for large $\Delta/T$ the probe limit becomes unreliable. However, we do expect qualitatively similar behavior of the (quantum) phase transition in the backreacted case \cite{inprogress}.

\begin{figure}
 \includegraphics[width=0.45\textwidth]{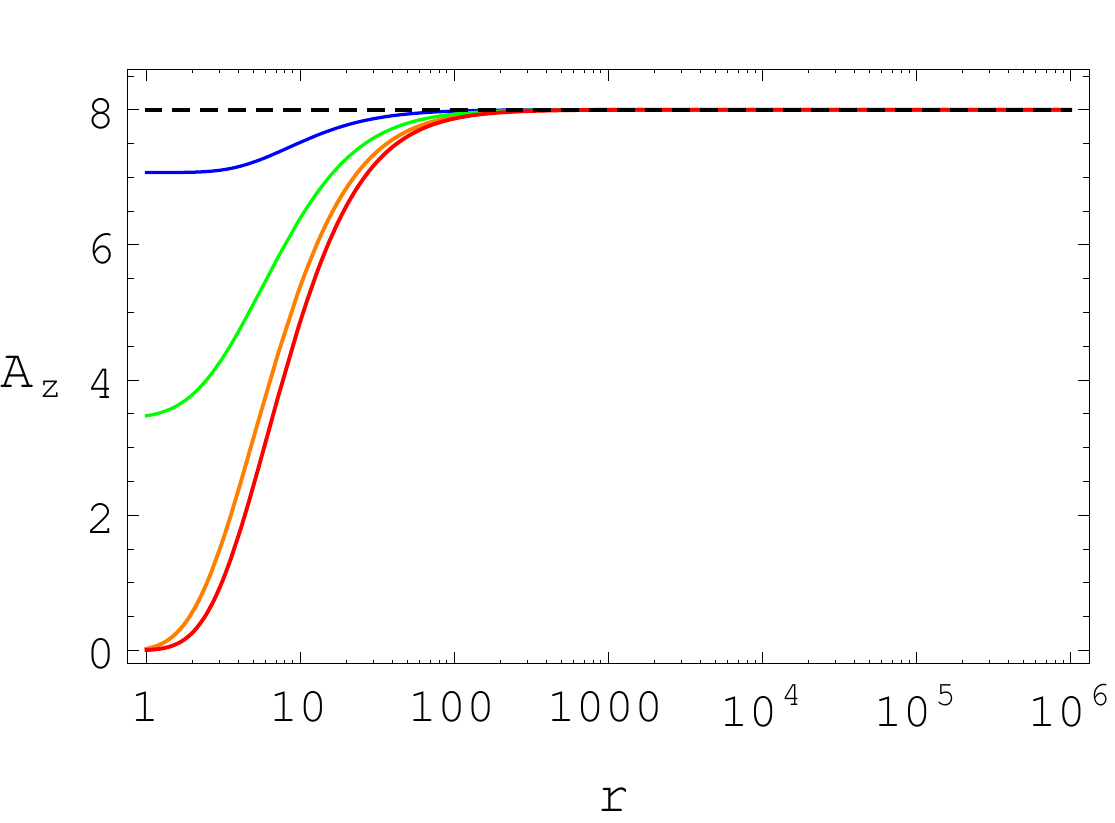}
 \includegraphics[width=0.45\textwidth]{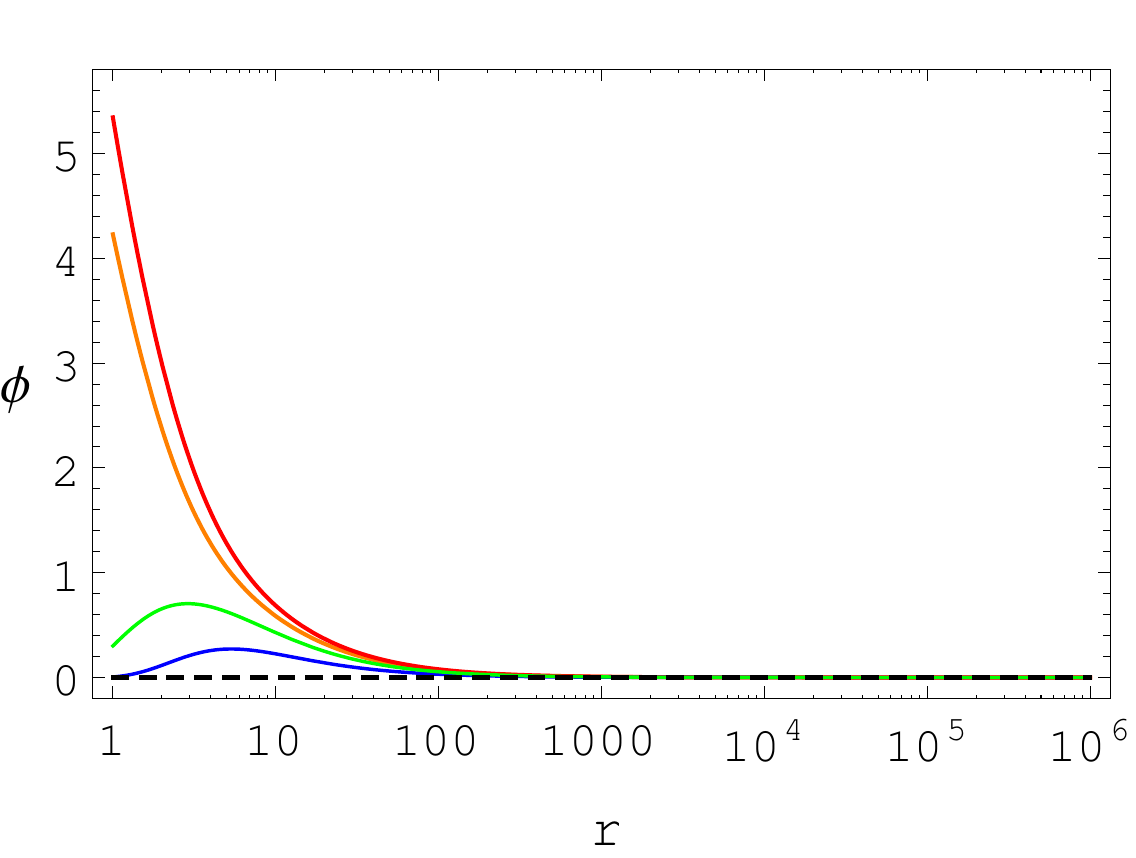}
\caption{\label{fig:azphirprofile} \small Holographic RG flow of the axial vector and scalar field for $\pi T/b=0.125$ 
and $M/b= 0.398$ (blue), $0.696$ (green), $0.875$ (orange), $0.995$ (red). }
\end{figure}

As we have argued the quantum phase transition in the quantum field theoretical model can be understood from the low energy perspective as a transition from the action (\ref{eq:lowenergychiral}) to (\ref{eq:lowenergygapped}). Both of them are special cases of (\ref{eq:lagrangian}) with the particular choices of parameters $M=0, b=b_\text{eff}$ for the ungapped and $M=\Delta, b=0$ for the gapped case. In fact our holographic model reproduces this low energy behavior with the limitations that arise by working at finite temperature. We remind the reader
that the holographic direction has to be understood as an energy scale. The profile of the functions $A_z^5(r)$ and $\phi(r)$ represent therefore
the running from high to low energies of the UV couplings $(M,b)$. We show typical profiles in Fig. \ref{fig:azphirprofile}.

In the topological phase ($b\gg M$) with significant Hall conductivity $A^5_z(r)$ stops running at some value $r>r_h$ and attains its IR value $A^5_z(r_h)$. This value determines the Hall conductivity and therefore is the holographic analogue of $b_\text{eff}$.
The scalar field first starts growing as one goes into the interior of AdS but turns around at a finite value of $r$ and becomes very small at the horizon. In the example of the plot for blue curve $\phi(r_h) \simeq 10^{-3}$. It does not vanish exactly since a small thermally induced gap is naturally expected to exists for all values of $M,b$.
Conversely, in the trivial phase the axial gauge field almost goes to zero at the horizon whereas the scalar field is monotonically increasing into the interior
of AdS until it hits the black hole horizon. 

\acknowledgements{
We thank A. Cortijo, Y. Ferreiro, D. Kharzeev, Y. W. Sun, M.A.H. Vozmediano for
useful discussions.
This work has been supported by project FPA2012-32828 and by the 
Centro de Excelencia Severo Ochoa Programme under grant SEV-2012-0249. }

\end{document}